\documentstyle[twoside,psfig]{article}

\catcode`\@=11
\long\def\@makefntext#1{
\protect\noindent \hbox to 3.2pt {\hskip-.9pt  
$^{{\eightrm\@thefnmark}}$\hfil}#1\hfill}		

\def\thefootnote{\fnsymbol{footnote}}
\def\@makefnmark{\hbox to 0pt{$^{\@thefnmark}$\hss}}	
	
\def\ps@myheadings{\let\@mkboth\@gobbletwo
\def\@oddhead{\hbox{}
\rightmark\hfil\eightrm\thepage}   
\def\@oddfoot{}\def\@evenhead{\eightrm\thepage\hfil
\leftmark\hbox{}}\def\@evenfoot{}
\def\sectionmark##1{}\def\subsectionmark##1{}}

\oddsidemargin=\evensidemargin
\renewcommand{\thefootnote}{\fnsymbol{footnote}}

\newcounter{sectionc}\newcounter{subsectionc}\newcounter{subsubsectionc}
\renewcommand{\section}[1] {\vspace{12pt}\addtocounter{sectionc}{1} 
\setcounter{subsectionc}{0}\setcounter{subsubsectionc}{0}\noindent 
	{\tenbf\thesectionc. #1}\par\vspace{5pt}}
\renewcommand{\subsection}[1] {\vspace{12pt}\addtocounter{subsectionc}{1} 
	\setcounter{subsubsectionc}{0}\noindent 
	{\bf\thesectionc.\thesubsectionc. {\kern1pt \bfit #1}}\par\vspace{5pt}}
\renewcommand{\subsubsection}[1] {\vspace{12pt}\addtocounter{subsubsectionc}{1}
	\noindent{\tenrm\thesectionc.\thesubsectionc.\thesubsubsectionc.
	{\kern1pt \tenit #1}}\par\vspace{5pt}}
\newcommand{\nonumsection}[1] {\vspace{12pt}\noindent{\tenbf #1}
	\par\vspace{5pt}}

\newcounter{appendixc}
\newcounter{subappendixc}[appendixc]
\newcounter{subsubappendixc}[subappendixc]
\renewcommand{\thesubappendixc}{\Alph{appendixc}.\arabic{subappendixc}}
\renewcommand{\thesubsubappendixc}
	{\Alph{appendixc}.\arabic{subappendixc}.\arabic{subsubappendixc}}

\renewcommand{\appendix}[1] {\vspace{12pt}
        \refstepcounter{appendixc}
        \setcounter{figure}{0}
        \setcounter{table}{0}
        \setcounter{lemma}{0}
        \setcounter{theorem}{0}
        \setcounter{corollary}{0}
        \setcounter{definition}{0}
        \setcounter{equation}{0}
        \renewcommand{\thefigure}{\Alph{appendixc}.\arabic{figure}}
        \renewcommand{\thetable}{\Alph{appendixc}.\arabic{table}}
        \renewcommand{\theappendixc}{\Alph{appendixc}}
        \renewcommand{\thelemma}{\Alph{appendixc}.\arabic{lemma}}
        \renewcommand{\thetheorem}{\Alph{appendixc}.\arabic{theorem}}
        \renewcommand{\thedefinition}{\Alph{appendixc}.\arabic{definition}}
        \renewcommand{\thecorollary}{\Alph{appendixc}.\arabic{corollary}}
        \renewcommand{\theequation}{\Alph{appendixc}.\arabic{equation}}
        \noindent{\tenbf Appendix \theappendixc #1}\par\vspace{5pt}}
\newcommand{\subappendix}[1] {\vspace{12pt}
        \refstepcounter{subappendixc}
        \noindent{\bf Appendix \thesubappendixc. {\kern1pt \bfit #1}}
	\par\vspace{5pt}}
\newcommand{\subsubappendix}[1] {\vspace{12pt}
        \refstepcounter{subsubappendixc}
        \noindent{\rm Appendix \thesubsubappendixc. {\kern1pt \tenit #1}}
	\par\vspace{5pt}}

\newcommand{\textlineskip}{\baselineskip=13pt}
\newcommand{\smalllineskip}{\baselineskip=10pt}

\def\eightcirc{
\begin{picture}(0,0)
\put(4.4,1.8){\circle{6.5}}
\end{picture}}
\def\eightcopyright{\eightcirc\kern2.7pt\hbox{\eightrm c}} 

\newcommand{\copyrightheading}[1]
	{\vspace*{-2.5cm}\smalllineskip{\flushleft
	{\footnotesize Report No. ANL-HEP-CP-00-094}\\
        {\footnotesize Presented at DPF2000, August 9-12, 2000.}}}

\def\abstracts#1#2#3{{
	\centering{\begin{minipage}{4.5in}\baselineskip=10pt\footnotesize
	\parindent=0pt #1\par 
	\parindent=15pt #2\par
	\parindent=15pt #3
	\end{minipage}}\par}} 

\newcommand{\bibit}{\nineit}

\renewenvironment{thebibliography}[1]
	{\frenchspacing
	 \ninerm\baselineskip=11pt
	 \begin{list}{\arabic{enumi}.}
	{\usecounter{enumi}\setlength{\parsep}{0pt}
	 \setlength{\leftmargin 12.7pt}{\rightmargin 0pt} 
	 \setlength{\itemsep}{0pt} \settowidth
	{\labelwidth}{#1.}\sloppy}}{\end{list}}

\newcounter{itemlistc}
\newcounter{romanlistc}
\newcounter{alphlistc}
\newcounter{arabiclistc}

\newcommand{\fcaption}[1]{
        \refstepcounter{figure}
        \setbox\@tempboxa = \hbox{\footnotesize Fig.~\thefigure. #1}
        \ifdim \wd\@tempboxa > 5in
           {\begin{center}
        \parbox{5in}{\footnotesize\smalllineskip Fig.~\thefigure. #1}
            \end{center}}
        \else
             {\begin{center}
             {\footnotesize Fig.~\thefigure. #1}
              \end{center}}
        \fi}

\newcommand{\tcaption}[1]{
        \refstepcounter{table}
        \setbox\@tempboxa = \hbox{\footnotesize Table~\thetable. #1}
        \ifdim \wd\@tempboxa > 5in
           {\begin{center}
        \parbox{5in}{\footnotesize\smalllineskip Table~\thetable. #1}
            \end{center}}
        \else
             {\begin{center}
             {\footnotesize Table~\thetable. #1}
              \end{center}}
        \fi}

\def\@citex[#1]#2{\if@filesw\immediate\write\@auxout
	{\string\citation{#2}}\fi
\def\@citea{}\@cite{\@for\@citeb:=#2\do
	{\@citea\def\@citea{,}\@ifundefined
	{b@\@citeb}{{\bf ?}\@warning
	{Citation `\@citeb' on page \thepage \space undefined}}
	{\csname b@\@citeb\endcsname}}}{#1}}

\newif\if@cghi
\def\cite{\@cghitrue\@ifnextchar [{\@tempswatrue
	\@citex}{\@tempswafalse\@citex[]}}
\def\citelow{\@cghifalse\@ifnextchar [{\@tempswatrue
	\@citex}{\@tempswafalse\@citex[]}}
\def\@cite#1#2{{$\null^{#1}$\if@tempswa\typeout
	{IJCGA warning: optional citation argument 
	ignored: `#2'} \fi}}

\def\pmb#1{\setbox0=\hbox{#1}
	\kern-.025em\copy0\kern-\wd0
	\kern.05em\copy0\kern-\wd0
	\kern-.025em\raise.0433em\box0}

\def\fnt#1#2{\footnotetext{\kern-.3em
	{$^{\mbox{\scriptsize #1}}$}{#2}}}

\def\fpage#1{\begingroup
\voffset=.3in
\thispagestyle{empty}\begin{table}[b]\centerline{\footnotesize #1}
	\end{table}\endgroup}

\def\runninghead#1#2{\pagestyle{myheadings}
\markboth{{\protect\footnotesize\it{\quad #1}}\hfill}
{\hfill{\protect\footnotesize\it{#2\quad}}}}
\font\tenrm=cmr10
\font\tenit=cmti10 
\font\tenbf=cmbx10
\font\bfit=cmbxti10 at 10pt
\font\ninerm=cmr9
\font\nineit=cmti9

\font\eightrm=cmr8

\def\qed{\hbox{${\vcenter{\vbox{			
   \hrule height 0.4pt\hbox{\vrule width 0.4pt height 6pt
   \kern5pt\vrule width 0.4pt}\hrule height 0.4pt}}}$}}

\renewcommand{\thefootnote}{\fnsymbol{footnote}}	

\begin{document}

\runninghead{Fully differential QCD corrections to single top $\ldots$}
{Fully differential QCD corrections to single top $\ldots$}

\normalsize\textlineskip
\thispagestyle{empty}
\setcounter{page}{1}

\copyrightheading{}			

\vspace*{0.88truein}

\fpage{1}
\centerline{\bf Fully differential QCD corrections to single top 
quark final states}
\vspace*{0.37truein}
\centerline{\footnotesize B. W. Harris\footnote{presenter; E-mail: 
harris@hep.anl.gov}$\:\:^1$, E. Laenen$^2$, L. Phaf$^2$, Z. Sullivan$^1$, 
S. Weinzierl$^2$}
\vspace*{0.05truein}
\centerline{\footnotesize\it $^1$ HEP Division, Argonne National Laboratory, 
Argonne, IL 60439, USA}
\vspace*{0.015truein}
\centerline{\footnotesize\it $^2$ NIKHEF Theory Group, Kruislaan 409, 1098 SJ 
Amsterdam, The Netherlands}

\vspace*{0.21truein}
\abstracts{A new next-to-leading order Monte Carlo program for calculation 
of fully differential single top quark final states is described and first 
results presented.  Both the $s$- and $t$-channel contributions are 
included.}{}{}


\vspace*{1pt}\textlineskip	
\section{Introduction}	        
\vspace*{-0.5pt}
\noindent

The electroweak production of single top quarks, despite being smaller 
in rate than it's strong interaction counterpart --- $t\bar{t}$ pair 
production, 
offers many physics possibilities for the Tevatron Run II and the LHC.
Direct measurement of $|V_{tb}|^2$ and sensitivity beyond the Standard Model 
are two examples \cite{1}.  It is also a background to some signals for 
the Higgs boson and various types of new physics.
Total rates for the $s$- and $t$-channel production processes 
have been calculated\cite{2,3} to next-to-leading 
order (NLO) in perturbative QCD.

Herein we give first results in our ongoing effort to calculate fully 
differential NLO QCD corrections to single top quark final 
states at hadron colliders.
The advantages of calculating cross sections to NLO   
in a fully differential manner are well known in general.
For example, only at NLO does a jet definition enter non-trivially.  
Additionally, systematic errors are always reduced when one can predict 
the cross in the experimentally visible region, as opposed to 
extrapolating the measured cross section to the full phase space 
to get a total rate.


\textheight=7.8truein
\setcounter{footnote}{0}
\renewcommand{\thefootnote}{\alph{footnote}}

\section{Method}
\noindent
The results presented below are from a 
calculation performed using the phase space slicing method \cite{4}.
Briefly, two cutoffs are used to isolate the regions containing 
soft and collinear singularities from the remainder 
of the three-body phase space.  The three-body squared matrix elements 
are integrated over these singular regions analytically.  
The soft singularities cancel upon addition of the 
virtual contributions, and the remaining collinear singularities are 
factorized into the parton distributions.  The integrations over 
the singularity-free portions of the three-body 
phase space are performed using Monte Carlo methods.  When all of 
the contributions are combined at the histogramming stage the various 
cutoff dependences cancel, since the cutoffs merely mark the 
boundary between the regions where the integrations were performed 
using analytic or Monte Carlo methods.

\section{Results}
\noindent

Using the method described in the previous section we have calculated the 
corrections to both $s$- and $t$-channel single-top-quark production.  Due to 
space limitations only $s$-channel results will be shown, and those only 
superficially.  Full details are forthcoming \cite{5}.

\begin{figure}
\centerline{\hbox{\psfig{figure=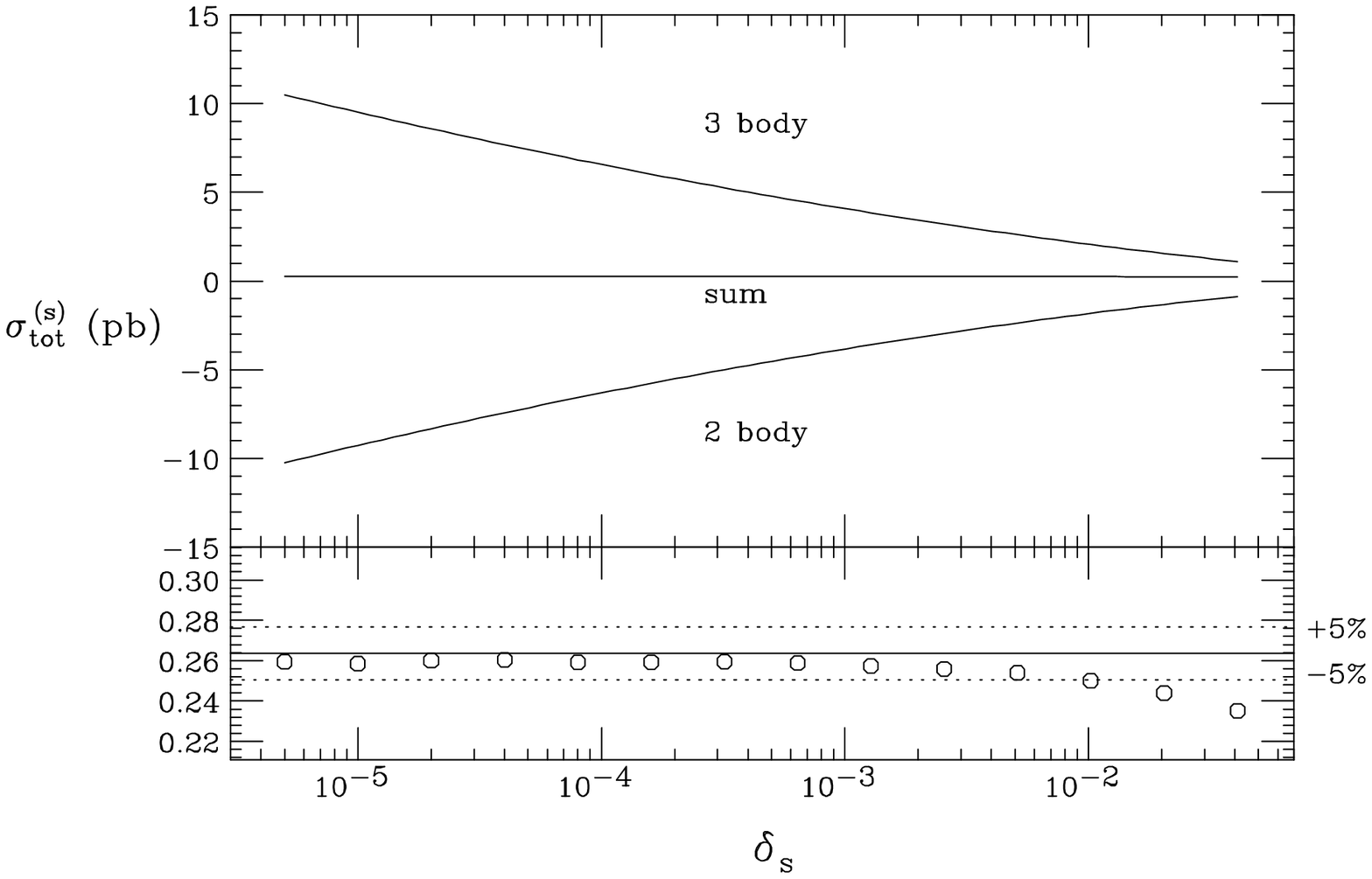,width=4.4in}}}
\fcaption{The next-to-leading order contribution to the single-top-quark 
total cross section via the $s$-channel process.  The two- and 
three-body contributions, together with their sum, are shown as a function 
of the soft cut-off $\delta_s$.  The bottom enlargement shows the sum 
(open circles) relative to $\pm 5 \%$ (dotted lines) of the analytic 
result (solid line).}
\label{fig:delta}
\end{figure}

A necessary check on the results based on this method is that infrared-safe 
physical observables are independent of the cut-offs used to delineate phase 
space, provided they are chosen small enough.  
The total cross section is an infrared-safe observable.
Shown in Fig.\ (\ref{fig:delta}) is the NLO 
{\em correction} to the total cross section as a function of the 
soft cut-off $\delta_s$ summed over $t$ and $\bar{t}$ final states 
for the $s$-channel production process compared with the known result\cite{2} 
using CTEQ4M\cite{6} parton distribution functions at a $\sqrt{S}=2$ TeV 
proton-antiproton machine.  The mass factorization and renormalization 
scales are taken to be $\mu_f=\mu_r=m_{\rm top}=175$ GeV.  The mass of 
the bottom quark was neglected and that of the $W$ gauge boson was 
taken to be $80.4$ GeV.  The Monte Carlo result for the NLO correction 
is within one percent of the analytical result of $0.264\; pb$ for 
$\delta_s < 10^{-3}$ implying better than a third of a percent agreement 
for the full NLO result of $0.922\; pb$.

\begin{figure}
\centerline{\hbox{
\psfig{figure=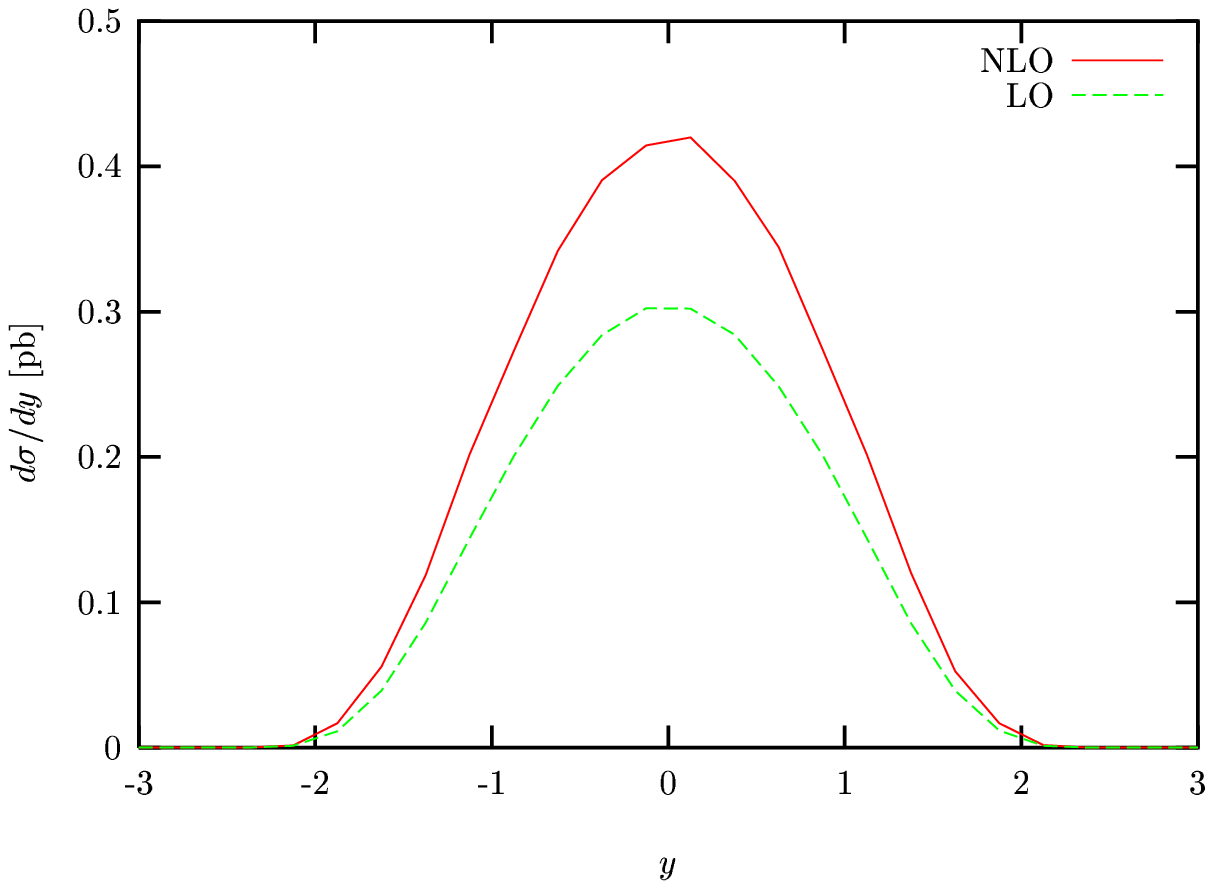,width=2.45in}
\psfig{figure=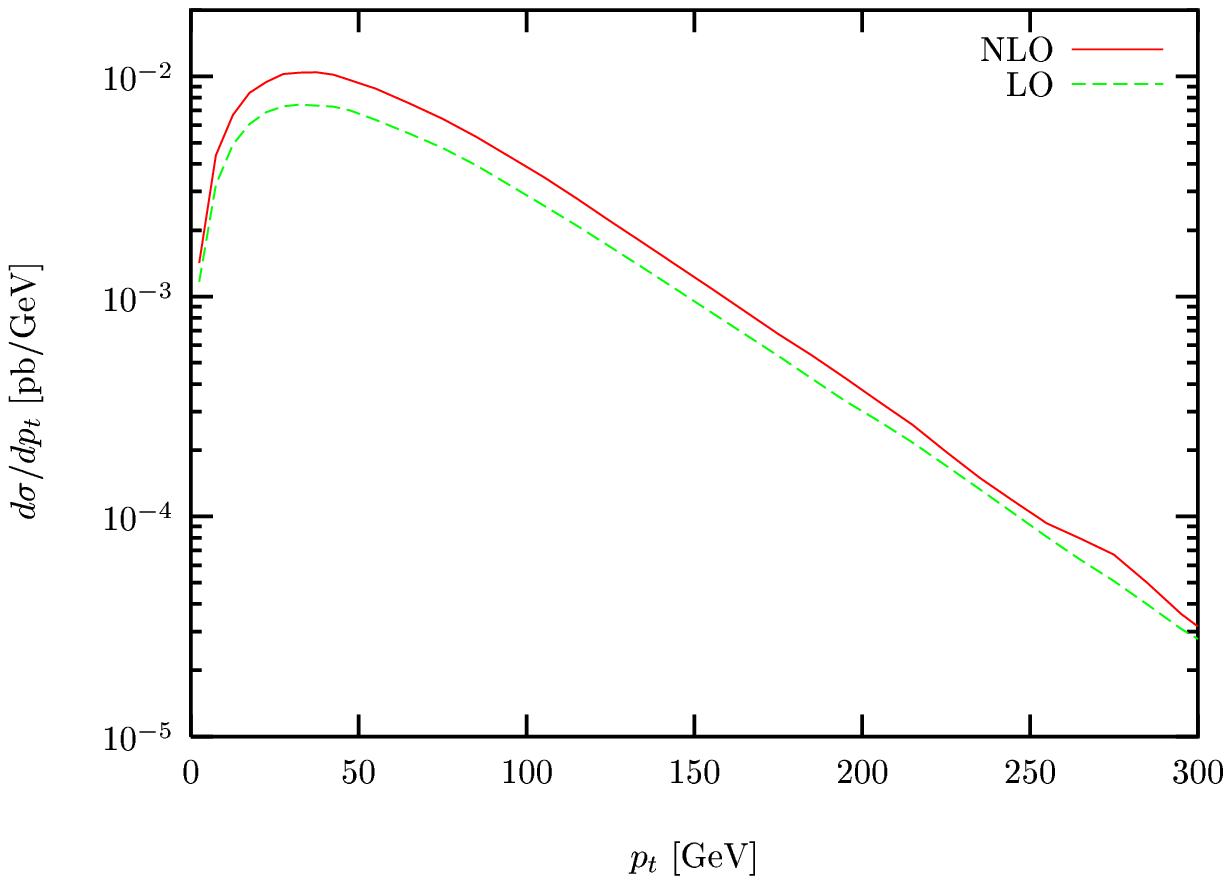,width=2.45in}}}
\fcaption{Leading (lower curves) and full next-to-leading (upper curves) order 
rapidity (left) and transverse momentum (right) spectra for the top 
quark produced via the $s$-channel process.}
\label{fig:pt}
\end{figure}

Shown in Fig.\ (\ref{fig:pt}) are the leading order (lower curves) and full 
next-to-leading order (upper curves) rapidity (left) and transverse momentum 
(right) distributions of the top quark produced via the $s$-channel 
process.  These curves were made using the same inputs as above, and no 
other cuts were applied.  These results are also 
cut-off independent, as expected.  Taking the ratio of NLO and LO the 
corrections are observed to be 
relatively independent of the rapidity and transverse momentum of the 
top quark, a somewhat atypical, but serendipitous (from the event generation 
point of view), situation that warrants further investigation.

In conclusion, we have performed a fully differential calculation of QCD 
corrections to electroweak production of single top quarks.  The method 
used allows for jet definitions and experimental cuts.  The corresponding 
computer code passes the necessary check of reproducing the previously 
known total rate.  First examination of the rapidity and transverse momentum 
of the top quark shows the corrections to be flat relative to leading order.  
Additional phenomenological studies are in order.
Further work is in progress on the corrections to $t-b-jet$ final states 
which will give a comprehensive set of tools for studying single-top-quark  
production at future hadron-hadron colliders.

\vspace*{0.2in}
\noindent
This work was supported in part by the United States Department of Energy, 
High Energy Physics Division, under contract W-31-109-Eng-38, and the
Foundation for Fundamental Research of Matter (FOM) and
the National Organization for Scientific Research (NWO), The Netherlands.

\nonumsection{References}

\end{document}